\documentclass[aps, prl, preprint, superscriptaddress, showpacs]{revtex4-1}

\usepackage{graphics}
\usepackage[english]{babel}
\usepackage{epstopdf}
\usepackage{amsmath}
\usepackage{amsfonts}
\usepackage{amssymb}
\usepackage[latin1]{inputenc} 
\usepackage[OT2,T1]{fontenc}

\begin{document}





\author{Simone Finizio}
\email{Corresponding Author: simone.finizio@psi.ch}
\affiliation{Swiss Light Source, Paul Scherrer Institut, 5232 Villigen PSI, Switzerland}

\author{Katharina Zeissler}
\affiliation{School of Physics and Astronomy, University of Leeds, Leeds LS2 9JT, United Kingdom}

\author{Sebastian Wintz}
\affiliation{Swiss Light Source, Paul Scherrer Institut, 5232 Villigen PSI, Switzerland}
\affiliation{Institute of Ion Beam Physics and Materials Research, Helmholtz-Zentrum Dresden-Rossendorf, 01328 Dresden, Germany}

\author{Sina Mayr}
\affiliation{Swiss Light Source, Paul Scherrer Institut, 5232 Villigen PSI, Switzerland}
\affiliation{Department of Materials, Laboratory for Mesoscopic Systems, ETH Z\"urich, 8093 Z\"urich, Switzerland}

\author{Teresa We{\ss}els}
\affiliation{Ernst Ruska-Centre for Microscopy and Spectroscopy with Electrons and Peter Gr\"unberg Institute 5, Forschungszentrum J\"{u}lich, 52425 J\"ulich, Germany}

\author{Alexandra J. Huxtable}
\affiliation{School of Physics and Astronomy, University of Leeds, Leeds LS2 9JT, United Kingdom}

\author{Gavin Burnell}
\affiliation{School of Physics and Astronomy, University of Leeds, Leeds LS2 9JT, United Kingdom}

\author{Christopher H. Marrows}
\affiliation{School of Physics and Astronomy, University of Leeds, Leeds LS2 9JT, United Kingdom}

\author{J\"org Raabe}
\affiliation{Swiss Light Source, Paul Scherrer Institut, 5232 Villigen PSI, Switzerland}


\title{Deterministic field-free skyrmion nucleation at a nano-engineered injector device}


\date{\today}


\begin{abstract}
\begin{bfseries}

Magnetic skyrmions are topological solitons that exhibit an increased stability against annihilation \cite{art:fert_topo_protection, art:buettner_skyrmion_energy_barrier}, and can be displaced with low current densities \cite{art:tomek_skyrmion_motion}, making them a promising candidate as an information carrier \cite{art:fert_topo_protection}. In order to demonstrate a viable skyrmion-based memory device, it is necessary to reliably and reproducibly nucleate, displace, detect, and delete the magnetic skyrmions. While the skyrmion displacement \cite{art:hofmann_skyrmion_hall_angle, art:kai_skyrmion_hall_angle, art:woo_skyrmion_motion_ferrimagnet, art:woo_skyrmion_motion} and detection \cite{art:zeissler_transport, art:maccariello_skyrmion_detection} have both been investigated in detail, much less attention has been dedicated to the study of the sub-ns dynamics of the skyrmion nucleation process. Only limited studies on the statics \cite{art:buettner_skyrmion_nucleation, art:jiang_blowing_skyrmions} and above-ns dynamics \cite{art:woo_skyrmion_nucleation} have been performed, leaving still many open questions on the dynamics of the nucleation process. Furthermore, the vast majority of the presently existing studies focus on the nucleation from random natural pinning sites \cite{art:buettner_skyrmion_nucleation,art:woo_skyrmion_nucleation}, or from patterned constrictions in the magnetic material itself \cite{art:buettner_skyrmion_nucleation, art:jiang_blowing_skyrmions}, which limit the functionality of the skyrmion-based device. Those limitations can be overcome by the fabrication of a dedicated injector device on top of the magnetic material \cite{art:hrabec_skyrmion_nucleation}. In this study, we investigate the nucleation of magnetic skyrmions from a dedicated nano-engineered injector, demonstrating the reliable magnetic skyrmion nucleation at the remnant state. The sub-ns dynamics of the skyrmion nucleation process were also investigated, allowing us to shine light on the physical processes driving the nucleation.

\end{bfseries}
\end{abstract}



\maketitle



Magnetic skyrmions are topological quasi-particles that can be stabilized in perpendicularly magnetized (PMA) materials exhibiting an anti-symmetric exchange interaction (Dzyaloshinskii-Moriya - DM - interaction). The DM interaction arises from the breaking of the inversion symmetry in the magnetic material \cite{art:dzyaloshinskii_DMI, art:moriya_DMI}. For multilayered PMA thin film systems where the inversion symmetry is broken at the interface between the different layers that compose the superlattice stack, the resulting interfacial DM interaction leads to the stabilization of N\'eel-type magnetic domain walls and skyrmions \cite{art:finizio_CIDWM, art:boulle_skyrmions, art:zeissler_transport}. Examples of multilayer stacks where room-temperature N\'eel-type magnetic skyrmions can be stabilized include Pt/Co/Ir \cite{art:moreau_luchaire_skyrmions_RT}, Pt/Co/MgO \cite{art:boulle_skyrmions}, and Pt/CoFeB/MgO \cite{art:kai_skyrmion_hall_angle}.

As a consequence of their non-trivial topology, magnetic skyrmions exhibit an increased stability against annihilation and pinning at defects \cite{art:fert_topo_protection, art:buettner_skyrmion_energy_barrier}, a topological contribution to the Hall resistivity \cite{art:nagaosa_skyrmion_topology, art:zeissler_transport}, and they can be displaced with low current densities down to 10$^6$-10$^7$ Am$^{-2}$ for single-crystalline materials exhibiting bulk DM interaction at cryogenic temperatures \cite{art:fert_topo_protection, art:tomek_skyrmion_motion}. The combination of these properties makes magnetic skyrmions particularly interesting both for fundamental studies and for applications in novel non-volatile magnetic memory concepts such as the skyrmion racetrack memory \cite{art:fert_topo_protection}.


In order to fabricate a viable skyrmion-based memory, the controllable and reproducible nucleation, motion, detection, and deletion of the magnetic skyrmions needs to be demonstrated. Skyrmion motion has been object of profound attention in the last years, leading to not only the demonstration of a reliable current-driven motion \cite{art:hofmann_skyrmion_hall_angle, art:kai_skyrmion_hall_angle, art:woo_skyrmion_motion_ferrimagnet, art:woo_skyrmion_motion}, but also to the study of the dynamics of the spin-orbit torque (SOT)-induced skyrmion motion \cite{art:kai_skyrmion_hall_angle}. Due to the influence of the skyrmion topology on its transport properties (giving rise e.g. to the topological Hall effect \cite{art:nagaosa_skyrmion_topology}), also the electrical detection of magnetic skyrmions has been thoroughly investigated, ranging from the investigation of the topological Hall effect in skyrmion crystals in materials exhibiting bulk DM interaction \cite{art:nagaosa_skyrmion_topology} to the detection of isolated magnetic skyrmions in multilayer stacks exhibiting interfacial DM interaction \cite{art:zeissler_transport, art:maccariello_skyrmion_detection}.


\begin{figure*}[!h]
 \includegraphics{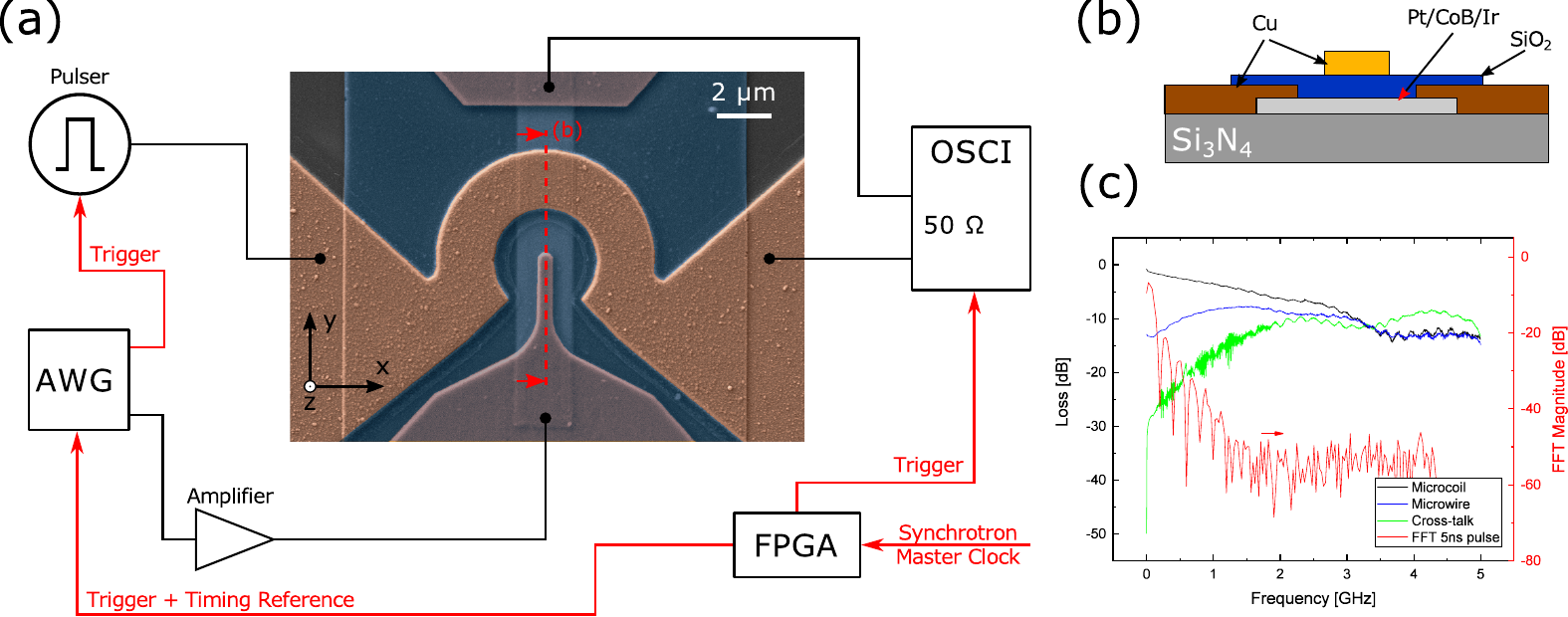}
 \caption{Schematic overview of the skyrmion nucleation device. (a) False color scanning electron micrograph of the injector structure employed for the skyrmion nucleation, with a schematic overview of the electronic setup employed for the pump-probe experiments. (b) Schematic cut view of the skyrmion injector sample (see red dashed line in (a)), showing the composition of the different layers. (c) S$_{21}$ parameters for the microwire and microcoil and cross-talk between the microwire and the microcoil. The fast-Fourier transform (FFT) of a 5 ns long pulse is also depicted, showing that the cutoff frequency for this excitation is on the order of 1 GHz, where the cross-talk between the microwire and the microcoil is much lower than the S$_{21}$ parameter for both components.}
 \label{fig:Sample_Setup}
\end{figure*}

However, only a limited experimental attention has been dedicated to the controlled nucleation of such magnetic quasi-particles. These studies mostly focus on the nucleation of magnetic skyrmions arising from natural or artificially fabricated defects in the magnetic material \cite{art:buettner_skyrmion_nucleation, art:woo_skyrmion_nucleation}, or from constrictions patterned directly into the magnetic material \cite{art:buettner_skyrmion_nucleation, art:jiang_blowing_skyrmions}. However, the properties of a natural defect site cannot be directly controlled, as different defect sites are not reproducibly equal both within the same and between different samples, making them unsuitable for the reliable and reproducible skyrmion nucleation that would be required in a device. Furthermore, due to the difficulty in determining the exact properties of the material in the defect site (e.g. the local values of the saturation magnetization or of the PMA), the interpretation of the processes occurring during the nucleation has to rely on \emph{a-priori} assumptions on the properties of the defect site \cite{art:woo_skyrmion_nucleation, art:buettner_skyrmion_nucleation}, or involve challenging investigations of the material \cite{art:hanneken_stm_dmi}. In the case that the skyrmion nucleation is achieved through a patterned constriction in the magnetic material \cite{art:buettner_skyrmion_nucleation, art:jiang_blowing_skyrmions}, the fabrication of the constriction itself requires a precise patterning, unnecessarily constraining its geometry. A constrained device geometry leads to a limited functionality of the magnetic device, complicating the design of a skyrmion-based memory employing such nucleators.

The fabrication of a constriction in the magnetic material leads to a non-uniform local distribution of the current density, which is then employed for the spatially-defined nucleation of the magnetic skyrmions \cite{art:buettner_skyrmion_nucleation, art:jiang_blowing_skyrmions}. However, one can also employ a patterned contact structure for achieving a non-uniform distribution of the current density, as proposed in the work of A. Hrabec \emph{et al.} \cite{art:hrabec_skyrmion_nucleation}. This design is a viable alternative for the electrical skyrmion nucleation, as a dedicated injector requires neither the direct patterning of the magnetic material nor to rely on natural or artificial defect sites. However, no detailed experimental investigations on the processes (and on their dynamics) that lead to the nucleation of a magnetic skyrmion from a dedicated injector device have been carried out as of now, leaving open questions on the physical mechanisms driving the nucleation.

Moreover, all of the currently available studies on the current-induced skyrmion nucleation require a permanently applied out-of-plane magnetic field for the stabilization of the magnetic skyrmions. The magnetic fields range in magnitude from sub-10 mT \cite{art:buettner_skyrmion_nucleation, art:jiang_blowing_skyrmions, art:hrabec_skyrmion_nucleation} to above 100 mT \cite{art:woo_skyrmion_nucleation}. The requirement of a permanently applied magnetic field is strongly detrimental for an application of magnetic skyrmions as an information carrier. A completely field-free stability of the nucleated skyrmions would instead be strongly desirable, as this would avoid the requirement of integrating permanent out-of-plane magnetic fields in the skyrmion-based device.


With the experiments presented in this work, we have overcome the issues discussed above by demonstrating the controlled skyrmion nucleation from a nano-engineered injector device. The injector was tailored to achieve a high current density at its tip, designating the region where the magnetic skyrmion is nucleated. The injector was fabricated on top of a Ta(3.2 nm)/Pt(2.6 nm)/[Co$_{68}$B$_{32}$(0.8 nm)/Ir(0.4 nm)/Pt(0.6nm)]$_{\times 3}$/Pt(2.1 nm) PMA stack. This stack, from now on referred to as Pt/Co$_{68}$B$_{32}$/Ir, was tailored to allow for the stabilization of magnetic skyrmions at the remnant state, therefore removing the requirement to carry out the experiments in presence of a permanently applied magnetic field. Additionally, the patterning of a nanostructured injector device on top of the Pt/Co$_{68}$B$_{32}$/Ir stack allows for the deterministic current-induced nucleation of magnetic skyrmions without the necessity to rely on random natural defects of the material, or on constrictions patterned directly into the material. Finally, thanks to the use of time-resolved scanning transmission X-ray microscopy (STXM), the sub-ns dynamical processes leading to the current-induced skyrmion nucleation (and magnetic field-induced deletion) could be unraveled.


\begin{figure}[!h]
 \includegraphics{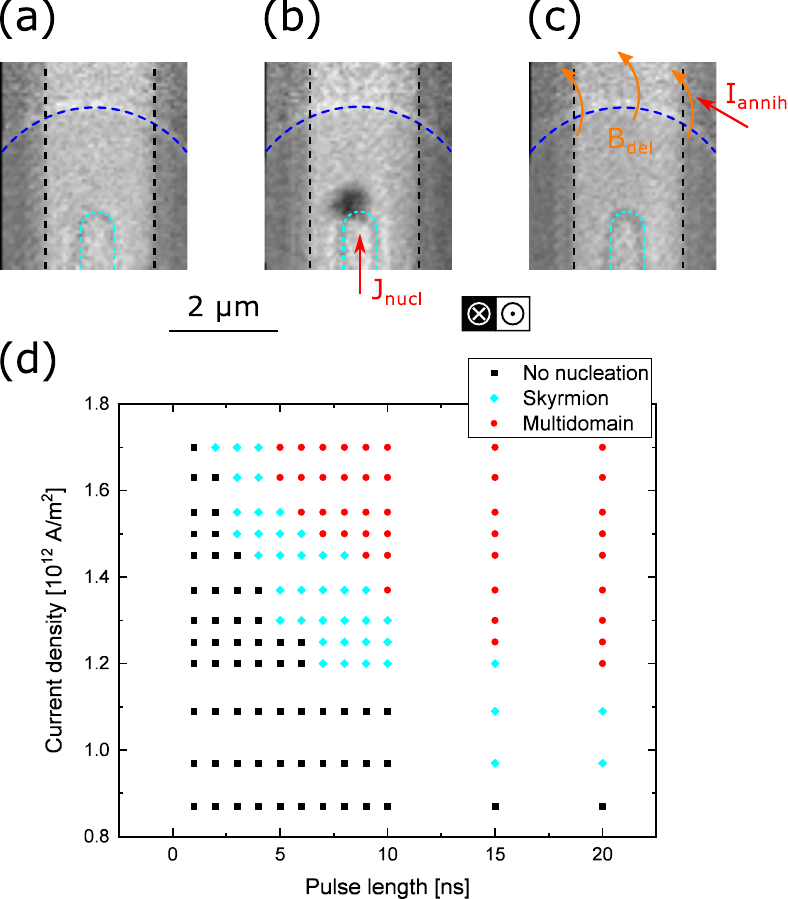}
 \caption{Quasi-static investigation of the skyrmion nucleation and deletion processes. (a-c) Quasi-static XMCD-STXM images of the current-induced nucleation and field-induced deletion of a magnetic skyrmion. (a) Initial uniform magnetic configuration. The edges of the microwire are marked by the black dashed lines, the edge of the injector is marked by the light blue dashed line, and the edge of the microcoil is marked by the dark blue dashed line. (b) Nucleation of a magnetic skyrmion (area of dark contrast in the image) by injecting a 5 ns long current pulse in the microwire. (c) Recovery of the initial magnetic configuration by injecting a current pulse across the microcoil, leading to the generation of an out-of-plane magnetic field pulse. (d) Dependence of the current density required to nucleate an isolated magnetic skyrmion on the duration of the current pulse injected from the skyrmion injector. A skyrmion nucleation window can be observed from these measurements. The current density shown in (d) was calculated in the region of the microwire where the current density exhibits a uniform distribution. All of the static XMCD-STXM images employed to obtain the results shown in (d) can be found in the supplementary information.}
 \label{fig:Static_Nucleation}
\end{figure}


For the skyrmion nucleation and deletion experiments reported here, we fabricated 500 nm wide Cu skyrmion injector structures on top of a Pt/Co$_{68}$B$_{32}$/Ir microwire (see Figs. \ref{fig:Sample_Setup}(a-b) and the Methods section). Thanks to the use of Cu for the fabrication of the electrical contacts, the magnetic configuration of the Pt/Co$_{68}$B$_{32}$/Ir microwire could be investigated also in the regions covered by the Cu contacts (see the Methods section). The skyrmion nucleation (starting from a uniformly magnetized state) is achieved by the injection of short current pulses from the Cu contact into the magnetic microwire. In order to reset the magnetic configuration back to a uniformly magnetized state within a ns timescale, a necessary condition to be able to perform time-resolved pump-probe experiments, an $\Omega$-shaped Cu microcoil was fabricated on top of the microwire. This allows for the generation of ns-wide out-of-plane magnetic field pulses that can be employed to magnetically saturate the region around the injector. Further details about the sample fabrication are provided in the Methods section. Note that the microwire and the microcoil are electrically insulated through the use of a 200 nm thick SiO$_2$ insulation layer, and that the geometry of the samples was optimized to minimize reflections of the injected current pulses, and to minimize the cross-talk between the microcoil and the microwire (see the Methods section and Fig. \ref{fig:Sample_Setup}(c) for more details).

The Pt/Co$_{68}$B$_{32}$/Ir multilayer stacks employed here stabilize N\'eel-type skyrmions at the remnant state (see Ref. \cite{art:finizio_CIDWM} for a proof that N\'eel-type domain walls are stabilized in this PMA stack). The zero-field skyrmion stability provides a strong motivation to use Pt/Co$_{68}$B$_{32}$/Ir stacks with respect to other common multilayer stacks that require the presence of an external out-of-plane magnetic field for the skyrmion stability \cite{art:moreau_luchaire_skyrmions_RT, art:buettner_gyration, art:zeissler_pinning, art:zeissler_transport, art:woo_skyrmion_motion_ferrimagnet, art:woo_skyrmion_nucleation, art:kai_skyrmion_hall_angle, art:hofmann_skyrmion_hall_angle, art:buettner_skyrmion_nucleation, art:jiang_blowing_skyrmions, art:hrabec_skyrmion_nucleation}. Because of this, we could perform the nucleation experiments at the remnant state, where the influence of external static magnetic fields on the nucleation and deletion processes can be safely ignored. This allows for an easier interpretation of the physical processes causing their nucleation and deletion, and removing the requirement to integrate a permanently-applied magnetic field with the skyrmion-based device.

\begin{figure}[!h]
 \includegraphics{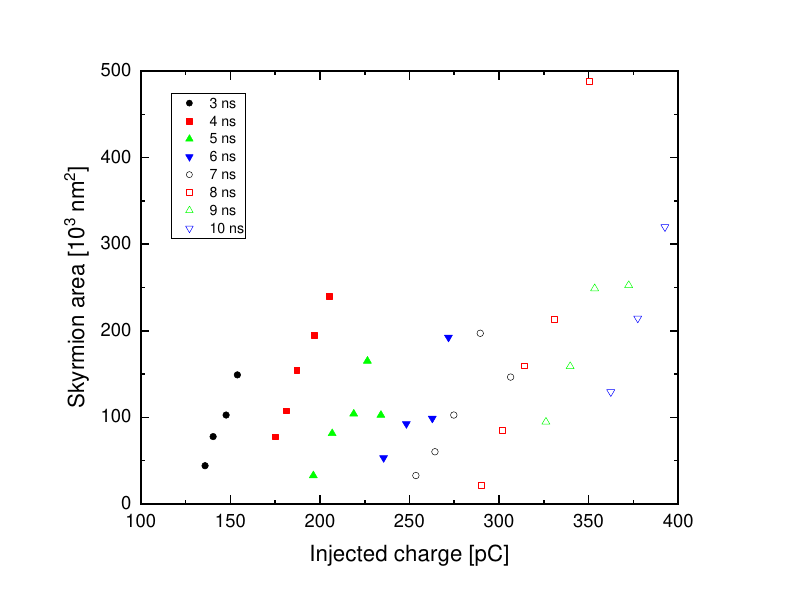}
 \caption{Dependence of the area of the magnetic skyrmions nucleated with a current pulse with respect to the charge injected in the Pt/Co$_{68}$B$_{32}$/Ir stack by the current pulse. Different pulse widths have been considered and, for a given pulse width, a roughly linear dependence of the area of the magnetic skyrmion with respect to the charge injected in the Pt/Co$_{68}$B$_{32}$/Ir stack can be observed.}
 \label{fig:Skyrmion_Size_vs_Injected_Charge}
\end{figure}


The skyrmion nucleation experiments were performed by STXM imaging both in the quasi-static and pump-probe regimes, using the X-ray magnetic circular dichroism (XMCD) effect \cite{art:schuetz_xmcd} as a means to obtain the magnetic contrast (see the Methods section for additional details about the technique). A first verification of the conditions necessary for the skyrmion nucleation at the nanostructured injector contact was performed by quasi-static STXM imaging. Here, the magnetic configuration after the injection of a single current pulse of variable widths and amplitudes (the current densities shown in this work were calculated in the section of the microwire exhibiting a uniform current density - see the supplementary information and the Methods section for additional details) was investigated. Prior to the injection of each pulse, the microwire was initialized in a uniformly magnetized state by injecting a 4 ns wide current pulse across the microcoil (at a peak current of 350 mA, resulting in a peak out-of-plane magnetic field of 50 mT in the region surrounding the injector - see the Methods section and the supplementary information for additional details). An example of the quasi-static XMCD-STXM imaging experiments carried out in this part of our work is shown in Fig. \ref{fig:Static_Nucleation}(a-c), also depicting the nucleation and deletion protocol employed for the time-resolved experiments presented in the sections below.

As summarized in Fig. \ref{fig:Static_Nucleation}(d), we observed three different scenarios depending on the width and amplitude of the injected current pulse (all of the images employed to obtain the results shown in Fig. \ref{fig:Static_Nucleation}(d) are shown in the supplementary information). For low current densities (black squares in Fig. \ref{fig:Static_Nucleation}(d)), the uniformly magnetized state remained unchanged upon the injection of the current pulse. For high current densities (red circles in Fig. \ref{fig:Static_Nucleation}(d)), a multidomain state was nucleated. Between these two cases, a window of pulse widths and amplitudes leading to the reproducible nucleation of an isolated magnetic skyrmion was observed (light blue diamonds in Fig. \ref{fig:Static_Nucleation}(d)). A monotonic decrease of the threshold current density for both the nucleation of an isolated skyrmion and for the stochastic nucleation of a multi-domain state with the width of the current pulse can also be noticed in Fig. \ref{fig:Static_Nucleation}(d). It is also noteworthy that for the pulse widths and amplitudes considered here it was possible to nucleate a magnetic skyrmion with a pulse energy down to about 500 pJ, which is comparable to the energy required to write a bit in commercial NOR flash memories \cite{art:wang_memory_energy}. Further optimizations in the energy efficiency of the skyrmion nucleator presented here could therefore lead to a better energy performance than NOR flash memories, proving that the encoding of digital bits as magnetic skyrmions can be a competitive alternative.

From the images where the nucleation of an isolated skyrmion was observed in Fig. \ref{fig:Static_Nucleation}(d), we extracted the area of the nucleated magnetic skyrmion as a function of the injected electrical charge. These results are shown in Fig. \ref{fig:Skyrmion_Size_vs_Injected_Charge}, where it is possible to observe that the area of the magnetic skyrmion exhibits a roughly linear dependence on the injected charge for a given pulse width, following a similar behavior observed for the partial SOT-induced switching of Pt/Co/AlO$_x$ nanodots \cite{art:baumgartner_switching}, hinting that the SOTs are one of the driving mechanisms behind the skyrmion nucleation process. For longer pulses, the amount of charge required for the nucleation of a magnetic skyrmion with a similar area is higher than for shorter pulses with higher current densities. This behavior, along with the observation that the minimum nucleation current density decreases with an increasing pulse width, indicates that part of the energy of the pulse is spent in the heating of the material. The results are in agreement with the currently employed models for SOT-assisted magnetization switching \cite{art:garello_SOTs, art:buettner_skyrmion_nucleation}, where a reduction in the switching current density with an increase of the temperature was observed \cite{art:jinnai_SOT_temperature}. A minimum skyrmion size with an area of 2.5 $\times$ 10$^4$ nm$^2$, corresponding to a diameter of about 50 nm, was observed from the results presented in Fig. \ref{fig:Skyrmion_Size_vs_Injected_Charge}. This minimum skyrmion size is compatible with calculations of the domain periodicity at the remnant state in similar PMA multilayer stacks optimized for a high interfacial DM interaction \cite{art:moreau_luchaire_skyrmions_RT}.



\begin{figure*}[!h]
 \includegraphics{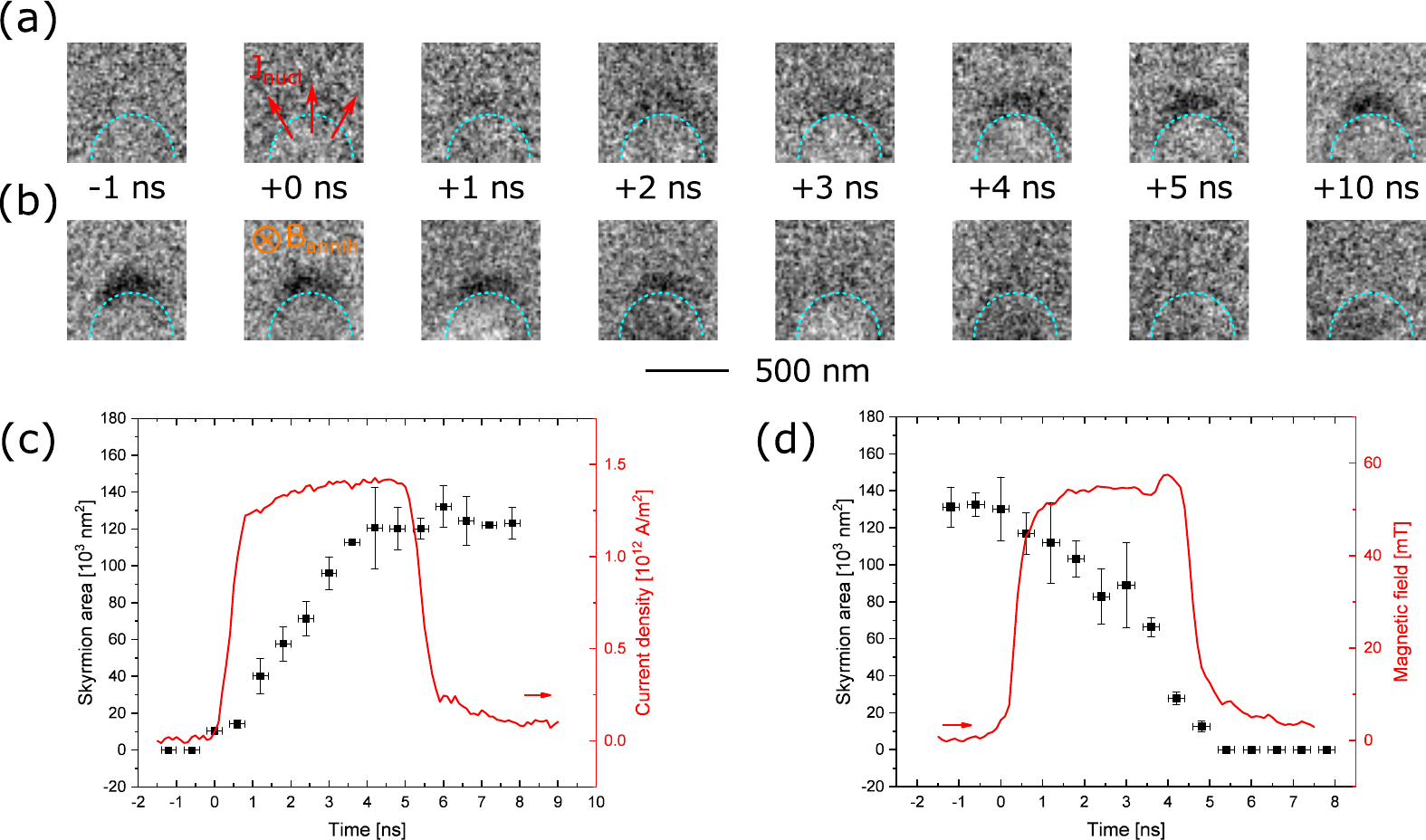}
 \caption{Dynamical imaging of the time-resolved current-induced skyrmion nucleation and magnetic field-induced skyrmion deletion processes. (a-b) Snapshots of differential time-resolved STXM images of (a) the nucleation and (b) the deletion processes. The edges of the injector are marked by the light blue dashed lines in the figure. (c) Time-resolved variation of the area of the magnetic skyrmion during the current-induced nucleation process. (d) Time-resolved variation of the area of the magnetic skyrmion during the field-induced skyrmion deletion process. The width of the magnetic field pulse was tailored to be equal to the time required for the complete deletion of the magnetic skyrmion. The magnitude of the magnetic field shown in (d) depicts the magnetic field at the tip of the injector structure determined by finite-element simulations (see the supplementary information and the Methods section).}
 \label{fig:Dynamic_Nucleation}
\end{figure*}

After investigating the quasi-static skyrmion nucleation (and deletion) processes, we turned our attention to the time-resolved imaging of the dynamical processes behind the current-induced skyrmion nucleation and the magnetic field-induced skyrmion deletion. These experiments were performed by time-resolved STXM imaging using a 5 ns wide nucleation pulse with a current density of $1.4 \times 10^{12}$ Am$^{-2}$, and a 350 mA, 4 ns wide, deletion pulse. The results of one such time-resolved investigation are depicted in Fig. \ref{fig:Dynamic_Nucleation}, where both the current-induced nucleation (Fig. \ref{fig:Dynamic_Nucleation}(a)) and the magnetic field-induced deletion (Fig. \ref{fig:Dynamic_Nucleation}(b)) are shown (videos of the time-resolved images can be found in the supplementary information). Figs. \ref{fig:Dynamic_Nucleation}(c-d) depict respectively the time-resolved variation of the skyrmion area during the nucleation and deletion. Due to the requirements of the pump-probe technique, the current-induced nucleation and magnetic field-induced annihilation processes were repeated about 10$^{10}$ times at a repetition frequency of 500 kHz, demonstrating that the nucleation and deletion processes are, within the limitations of the pump-probe technique, both reproducible and deterministic.

Thanks to the possibility of precisely determining the time position of the zero delay between pump and probe, the time-resolved variation of the skyrmion area can be overlaid and compared with the current/magnetic field excitation (see the Methods section for additional details). For both nucleation and deletion, it can be immediately observed from Figs. \ref{fig:Dynamic_Nucleation}(c-d) that they do not exhibit a detectable incubation time within the 200 ps time step employed here and therefore start synchronously with the onset of the current/magnetic field pulse. The absence of an incubation time for the nucleation process provides, after the quasi-static behavior shown in Fig. \ref{fig:Skyrmion_Size_vs_Injected_Charge}, another indication that the nucleation is driven by SOT-induced switching of the magnetization, which instantaneously responds to the excitation \cite{art:baumgartner_switching}.


Another observation that can be made from the time-resolved variation of the skyrmion area shown in Figs. \ref{fig:Dynamic_Nucleation}(c-d) is that, for both the nucleation and deletion processes, the area of the magnetic skyrmion varies roughly linearly with the duration of the current/magnetic field pulse. This behavior is in agreement with the quasi-static experiments shown in Fig. \ref{fig:Skyrmion_Size_vs_Injected_Charge}, and additionally with time-resolved observations on the partial switching of Pt/Co/AlO$_x$ nanodots \cite{art:baumgartner_switching}, providing yet another indication that SOTs drive the current-induced skyrmion nucleation.



\begin{figure}[!h]
 \includegraphics{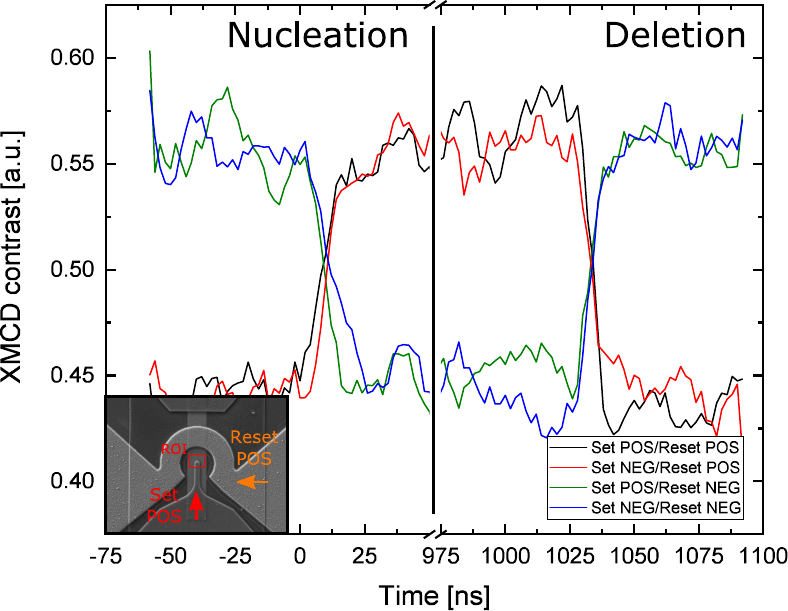}
 \caption{Skyrmion nucleation and deletion dynamics as a function of the direction of the nucleation current. Time-resolved variation in the XMCD contrast for the current-induced skyrmion nucleation and magnetic field-induced skyrmion deletion processes for different directions of the nucleation current and deletion field. A magnetic skyrmion can be nucleated independently from the direction of the nucleation current and from the orientation of the starting single-domain configuration (which is selected by the sign of the resetting magnetic field pulse). The inset shows the sign convention used for the nucleating current and deleting magnetic field, and the region of interest (ROI) employed for the calculation of the time-resolved XMCD contrast variation.}
 \label{fig:Directionality_Nucleation}
\end{figure}

We also investigated the influence of the direction of the electrical current employed for the nucleation of the magnetic skyrmions. Here, we performed pump-probe imaging experiments of both the nucleation and deletion processes with a 2 ns probing time step as a function of the direction of both the nucleating and deleting pulses. The time-resolved change of the magnetic contrast in the region surrounding the injector structure during the nucleation and deletion processes was measured. Those results are shown in Fig. \ref{fig:Directionality_Nucleation}. The sign convention for the direction of the nucleation current and deletion magnetic field is shown in the inset of the figure, while the area where the time-resolved change in the magnetic contrast was measured is marked by ROI in the same inset. From the time traces shown in Fig. \ref{fig:Directionality_Nucleation}, it is possible to observe that a magnetic skyrmion can be nucleated for all four possible combinations of the directions of the nucleating current and deleting magnetic field. Moreover, the dynamics of the nucleation and deletion processes occur on the same timescales in all four combinations, indicating that the nucleation process occurs independently from the direction of the current flowing in the injector.





In the absence of static external magnetic fields, the four possible combinations of the directions of the nucleation current and deletion field are symmetric. The magnitude of the torque generated by the SOT effect will therefore be equal for all four possible combinations \cite{art:baumgartner_switching}. If the injected spin current is sufficiently high, a switching of the magnetization will be observed \cite{art:garello_SOTs, art:baumgartner_switching, art:buettner_skyrmion_nucleation}, independently of the direction of the spin current. The sign of the torque will be affected by the orientation of the electrical current, but it will nonetheless lead to a switching of the magnetization \cite{art:baumgartner_switching}, in agreement with the results presented in Fig. \ref{fig:Directionality_Nucleation}.


Consequently, the results shown above allow us to conclude that the current-induced skyrmion nucleation process is driven by the SOT-induced switching of the magnetization in the area at the tip of the nano-engineered injector structure. It is however necessary to consider that the injection of the 5 ns-wide pulse considered here will lead, due to Ohmic losses, to the heating of the Pt/Co$_{68}$B$_{32}$/Ir microwire (see the supplementary information for a determination of the sample temperature from the time-resolved STXM images). The heating assists in the reduction of the critical current densities required to achieve a SOT-induced switching \cite{art:jinnai_SOT_temperature}. This indicates that the switching process is driven, for these pulse widths, by the synergy between the SOTs and the heating caused by the injected pulse. 

As shown in Figs. \ref{fig:Static_Nucleation} and \ref{fig:Dynamic_Nucleation} and in the figures and videos provided in the supplementary information, the nucleation of the magnetic skyrmion takes place at the tip of the injector structure. Nucleations outside of this region occur only for the pulses with current densities much higher than the single skyrmion nucleation threshold (see Fig. \ref{fig:Static_Nucleation}(d) and the images in the supplementary information). This can be explained by considering the geometry of the injector, and by the distribution of the current density during the injection of a current pulse (see the supplementary information for a finite-element simulation of the current density in the microwire). In particular, it can be observed that the magnitude of the current density at the tip of the injector is of about a factor 3 larger (in a circular region of about 150 nm in diameter) than the current density in the uniform section of the Pt/Co$_{68}$B$_{32}$/Ir microwire. The injector geometry presented here provides thus a localized "hot spot" with a higher current density that locally fulfills the requirements for the SOT-induced switching of the magnetization. On a first glance, this appears to be a similar concept to the SOT-induced nucleation of magnetic skyrmions at natural defects \cite{art:woo_skyrmion_nucleation, art:buettner_skyrmion_nucleation} or, especially, at constrictions patterned directly on the magnetic material, which rely on a localized non-uniformity in the current distribution to facilitate the skyrmion nucleation \cite{art:buettner_skyrmion_nucleation, art:jiang_blowing_skyrmions}. However, while some similarities can be observed, our approach exhibits the advantage that the skyrmion nucleation does not rely on any natural (or artificial) defects and constrictions on the magnetic material, but on the patterning of an injector structure on top of the magnetic material. This avoids the necessity for the fabrication of controlled pinning sites on the magnetic material and, with the proposed design, the skyrmion injector structure can be fabricated directly on top of a skyrmion racetrack, where the nucleated skyrmions can then be displaced by a current pulse injected across the microwire (whilst maintaining the injector structure electrically floating). This design provides a simple method for nucleating the magnetic skyrmions, while allowing for the engineering of the functionality of the material by tailoring the design of the electrical contacts to the required process (e.g. displacement of the magnetic skyrmions in a racetrack memory element). Furthermore, the zero-field skyrmion stability provides a framework where magnetic skyrmions can be manipulated, allowing for the possibility to fabricate skyrmion-based devices free of any external permanent magnets.



In conclusion, we have shown that the fabrication of a nano-engineered injector structure on top of a Pt/Co$_{68}$B$_{32}$/Ir superlattice allows for the deterministic nucleation of isolated magnetic skyrmions at the remnant state by the injection of ns-wide current pulses. Quasi-static and time-resolved STXM imaging of the skyrmion nucleation process allowed us to observe that the process is driven by the SOT-induced local switching of the magnetization, and quasi-static STXM investigations of the properties of the switching pulses (width and current density) allowed us to conclude that the heating of the microwire also plays an important role in the switching process by reducing the threshold current density for longer pulses. The combination of the flexibility of the geometry of the magnetic material given by the use of an injector structure for the skyrmion nucleation with the zero field stability of the magnetic skyrmions in the Pt/Co$_{68}$B$_{32}$/Ir superlattice stack offers a leap forward towards the device applicability of magnetic skyrmions. In particular, our approach permits the generation of a skyrmion "where and when we want it", and will provide an easy framework where to investigate the skyrmion properties in absence of external fields.




\section{Methods}

\subsection{Sample fabrication and characterization}
The microstructured wires were fabricated to a width of 2 $\mu$m by electron beam lithography followed by liftoff out of a Ta(3.2 nm)/Pt(2.6 nm)/[Co$_{68}$B$_{32}$(0.8 nm)/Ir(0.4 nm)/Pt(0.6nm)]$_{\times 3}$/Pt(2.1 nm) multilayer stack. The Pt/Co$_{68}$B$_{32}$/Ir stacks were deposited by magnetron sputtering at a rate of approximately 1 \AA s$^{-1}$ using an Ar plasma at a pressure of 3.2 mbar (base pressure of the deposition chamber in the low 10$^{-8}$ mbar range). More details on the deposition of a similar multilayer stack can be found in Ref. \cite{art:zeissler_pinning}. For the lithographical fabrication, a bilayer of methyl-methacrylate (MMA) and of poly(methyl-methacrylate) (PMMA) resist was spincoated on top of 200 nm thick x-ray transparent Si$_3$N$_4$ membranes on 200 $\mu$m thick high resistivity Si frames. The MMA/PMMA bilayer was exposed by electron beam lithography using a 100 kV Vistec EBPG 5000Plus electron beam writer with a writing dose of 1600 $\mu$C cm$^{-2}$. The exposed resist was developed by immersion for 90 s in a solution of methyl-isobutyl-ketone and isopropyl alcohol 1:3 in volume, followed by immersion in pure isopropyl alcohol for 60 s. After the deposition of the magnetic material, the unexposed resist, along with the magnetic film on top, was removed by immersion in pure acetone.

The microwires were contacted by 200 nm thick Cu electrodes fabricated by electron-beam lithography followed by liftoff. The Cu was deposited by thermal evaporation using a Balzers BAE250 thermal evaporator. One of the electrodes was fabricated in a rounded geometry with a 250 nm radius of curvature, designed to obtain a diverging current flow. To generate the magnetic field pulses employed for the deletion of the magnetic skyrmion, an $\Omega$-shaped, 400 nm thick, Cu microcoil was fabricated on top of the microwire once again by electron-beam lithography followed by a lift-off process. The microwire and the microcoil were electrically insulated through a 200 nm thick SiO$_2$ layer deposited between the two layers by electron beam evaporation (using a Univex 450 electron beam evaporator). For all the layers, injector, insulator, and microcoil, the same process employed for the patterning of the magnetic material was utilized. Cu was chosen as electrode material as this guarantees a high x-ray transmittivity at the Co L$_3$ edge, allowing for the investigation of the magnetic configuration of the sample also below the electrodes.

The quality of the samples fabricated according to the protocol described above was verified by scanning electron microscopy imaging (see Fig. \ref{fig:Sample_Setup}(a)). Their electrical properties (in particular, the impedance matching of both the Pt/Co$_{68}$B$_{32}$/Ir microwire and the microcoil, and the cross-talk between the two) were verified by time-domain reflectometry measurements using an Agilent Infiniium DCA-J 86100C sampling oscilloscope. With this technique, the S$_{21}$ parameters of the microwire, microcoil, and the cross-talk between them was determined. The results of this characterization are shown in Fig. \ref{fig:Sample_Setup}(c). The electrical resistance of the samples, employed for monitoring the quality of the sample during the quasi-static and time-resolved STXM investigations, was measured using a Keithley 2400 source meter, employing a probing current of 10 $\mu$A.

The magnetic properties of the Pt/Co$_{68}$B$_{32}$/Ir multilayer stacks were characterized by magnetometry measurements using a superconducting quantum interference device (SQUID). From these measurements, the temperature dependence of the saturation magnetization, employed to determine the time-resolved variation of the temperature close to the injector structure, could be determined.

\subsection{Skyrmion nucleation and deletion protocol}
The skyrmions were nucleated by injecting a current pulse across the injector structure. For the quasi-static investigations, the width and amplitude of the current pulse were varied from 1 to 20 ns and from $8.5 \times 10^{11}$ to $1.7 \times 10^{12}$ Am$^{-2}$ respectively. For the time-resolved investigations, 5 ns wide current pulses with a peak current density of $1.4 \times 10^{12}$ Am$^{-2}$ were employed. The current pulses were generated using a Tektronix AWG 7122C 10 GSa s$^{-1}$ arbitrary waveform generator (AWG) combined with a MiniCircuits ZPUL-30P non-inverting pulse amplifier. For the generation of the magnetic fields employed for the skyrmion deletion, 4 ns wide current pulses with a peak current of 350 mA were injected across the microcoil structure. A temporal gap of about 1 $\mu$s between the current and magnetic field pulses was employed. The magnetic field pulses were generated using an Avtech AVM-4-C 20 V pulse generator, which was triggered by a synchronization signal generated with the second channel of the AWG. Both the nucleation and deletion pulses were monitored using a Keysight DSO-S 404A 20 GSa s$^{-1}$real time oscilloscope. The oscilloscope was terminated with a 50 $\Omega$ impedance, allowing for the measurement of the current flowing across the microwire and the microcoil. The electrical connections to the sample are shown in Fig. \ref{fig:Sample_Setup}(a).

\subsection{STXM imaging}
The quasi-static and time-resolved STXM imaging of the skyrmion nucleation and deletion processes was performed using the STXM installed at the PolLux (X07DA) endstation of the Swiss Light Source \cite{art:pollux}. With STXM imaging, circularly-polarized monochromatic x-rays generated by the synchrotron light source (tuned to the Co L$_3$ absorption edge) are focused on the sample using diffractive optics (Fresnel zone plate). The x-rays transmitted through the sample are recorded using an avalanche photodiode (APD) as photon detector. To form an image, the sample is raster scanned with a piezoelectric stage. For the experiments presented here, a Fresnel zone plate with a 25 nm outermost zone width was employed. The width of the entrance and exit slits to the beamline monochromator was selected to achieve an x-ray beam spot on the order of 25-30 nm, defining the spatial resolution of the images presented in this work. 

The time-resolved images were acquired in the pump-probe regime illuminating the sample with circularly-polarized x-rays with negative helicity. Thanks to the use of a broadband APD, combined with a dedicated field-programmable gate array (FPGA) setup (which handles also the synchronization between the x-ray pulses and the electronics that excite the sample - see Fig. \ref{fig:Sample_Setup}(a)), the entire filling pattern of the synchrotron light source can be employed for the acquisition of the time-resolved images \cite{art:puzic_TR_STXM, art:finizio_TR_STXM}. Time steps of 2 ns and 200 ps were employed for the time-resolved images presented here. The synchronization between the pump and probe signals (in particular, the determination of the time position of the zero delay - or $t_0$) was determined using a laser diode with a rise time faster than 200 ps connected to the same electronic setup employed for the experiments. This allowed for a determination of $t_0$ within an accuracy of below 200 ps.

The time-resolved images shown in Fig. \ref{fig:Dynamic_Nucleation} and in the videos provided in the supplementary information were acquired as differential images. Here, the average magnetization configuration before the nucleation or after the deletion of the skyrmion (i.e. when the area around the magnetic wire is in a uniformly magnetized state) was subtracted from each frame of the time-resolved series. Similarly to an XMCD image, the resulting differential image shows the nucleation and deletion of the magnetic skyrmion through a variation in the image intensity.

The area of the magnetic skyrmions shown in Fig. \ref{fig:Skyrmion_Size_vs_Injected_Charge} and the time-dependent area of the magnetic skyrmion shown in Fig. \ref{fig:Dynamic_Nucleation} were determined through a thresholding of the quasi-static and time-resolved STXM images. The threshold was selected to be half of the width of the change in contrast between the skyrmion and the outlying domain of opposite magnetization (i.e. selecting the region corresponding to m$_z$ = 0 as threshold). To improve the statistics on the time-resolved images, a 3 point moving average in the time domain was employed.

\subsection{Finite-element simulations}
The current density distribution during the injection of a current pulse, as well as the spatial distribution of the magnetic field generated by injecting a pulse across the microcoil was simulated using the commercial multiphysics finite-elements simulation software ANSYS-Maxwell. For the simulations, the measured values of the current flowing through the Pt/Co$_{68}$B$_{32}$/Ir microwire or through the microcoil were employed. The results of the finite-element simulations performed for the current and magnetic field distributions are shown in the supplementary information. The values of the current density shown in Figs. \ref{fig:Static_Nucleation} and \ref{fig:Dynamic_Nucleation} were taken from the region of the microwire where the current density exhibits a uniform spatial distribution. Close to the injector structure, a strong spatial variation of the current density was observed, with a peak current density of about a factor 3 higher than in the uniform section.

\subsection{Determination of the sample temperature}
The temperature of the sample during the injection of the current pulse was determined from the time-resolved STXM images. In particular, the variation of the magnetic contrast in the uniformly magnetized region of the Pt/Co$_{68}$B$_{32}$/Ir microwire was measured. From this measurement, the time-resolved variation of the saturation magnetization of the sample (under the assumption that the uniformly magnetized region of the Pt/Co$_{68}$B$_{32}$/Ir microwire maintains its PMA) could be determined. The time-resolved variation of the sample temperature, shown in the supplementary information along with further details on the determination of the sample temperature, could then be determined by comparing the value of the saturation magnetization with quasi-static magnetometry measurements of the saturation magnetization as a function of the sample temperature.

\section{Acknowledgments}
\begin{acknowledgments}

This work was performed at the PolLux (X07DA) endstation of the Swiss Light Source, Paul Scherrer Institut, Villigen PSI, Switzerland. The research leading to these results has received funding from the European Community's Seventh Framework Programme (FP7/2007-2013) under grant agreement No. 290605 (PSI-FELLOW/COFUND), the Swiss National Science Foundation under grant agreement No. 200021\_172517, and the European Union's Horizon 2020 Projects MAGicSky (Grant No. 665095) and NFFA (Grant No. 654360), having benefited from the access provided by the Paul Scherrer Institut in Villigen PSI within the framework of the NFFA-Europe Transnational Access Activity. The authors thank P. Gambardella for helpful discussions.

\end{acknowledgments}

\end{document}